 \documentstyle[12pt]{article}
 \newcommand{\be}{\begin{equation}}
 \newcommand{\ee}{\end{equation}}
 \newcommand{\ba}{\begin{equation}\begin{array}{lll}}
 \newcommand{\ea}{\end{array}\end{equation}}
 \def\entry#1#2{\vbox{\hbox to 92truept{\hrulefill}\break
                \hbox{\vrule\vbox to 30truept{
                \vfill
                \hbox to 92truept{\hfill\quad\small #1\quad\hfill}\break
                \vfill
                \hbox to 92truept{\hfill\quad\small #2\quad\hfill}
                \break\vfill
                \hbox to 92truept{\hrulefill}}\vrule}}}

 \def\arrwv#1#2{\vbox to 68truept{\vfill
                \hbox to 92truept{\put(48,0){\line(0,-1){20}}\hfill}\break
                \vfill
                \hbox to 92truept{\hfill\small #1\hfill}\break
                \vfill
                \hbox to 92truept{\hfill\small #2\hfill}\break
                \hbox to 92truept{%
                \put(48,0){\vector(0,-1){20}}\hfill}\break
                \vfill}}

 \def\arrwh#1#2{\vbox to 30truept{\vfill
                \hbox to 92truept{\small\hfill #1\hfill}\break
                \hbox to 92truept{\rightarrowfill}\break
                \hbox to 92truept{\small\hfill #2\hfill}\break\vfill}}

 \def\d{\partial}
 \def\real{{\vrule height 1.0ex
             width 0.05em depth 0ex \kern -0.06em {\rm R}}}
 \def\Journal#1#2#3#4{{#1} {\bf #2}, #3 (#4)}

 \def\ANP{\em Ann.\ Physics (N.Y.)}
 \def\CQG{\em Class.\ Quantum Grav.}
 
 \def\IJMPA{{\em Int.\ J.\ Mod.\ Phys.} A}
 \def\IJMPD{{\em Int.\ J.\ Mod.\ Phys.} D}
 \def\MPLA{{\em Mod.\ Phys.\ Lett.} A}
 
 \def\NPBP{{\em Nucl. Phys.} (Proc.\ Suppl.)}
 \def\PLB{{\em Phys. Lett.}  B}
 
 \def\PRD{{\em Phys. Rev.} D}
 \def\TMP{\em Theor.\ Math.\ Phys.}
\begin{document}
\title{Geometrodynamical Formulation of Two-Dimensional Dilaton Gravity}
\author{Marco Cavagli\`a\thanks{E-mail:
cavaglia@aei-potsdam.mpg.de; web page: 
http://www.aei-potsdam.mpg.de/\~{}cavaglia\hfill}\\
\sl Max-Planck-Institut f\"ur Gravitationsphysik\\
\sl Albert-Einstein-Institut\\
\sl Schlaatzweg 1, D-14473 Potsdam, Germany}
\date{\today}
\maketitle
\begin{abstract}
Two-dimensional matterless dilaton gravity with arbitrary dilatonic
potential can be discussed in a unitary way, both in the Lagrangian and
canonical frameworks, by introducing suitable field redefinitions. The new
fields are directly related to the original spacetime geometry and in the
canonical picture they generalize the well-known geometrodynamical
variables used in the discussion of the Schwarzschild black hole. So the
model can be quantized using the techniques developed for the latter case.
The resulting quantum theory exhibits the Birkhoff theorem at the quantum
level.
%

\bigskip
\noindent
Pac(s) numbers: 04.20.Fy, 04.60.Ds, 04.60.Kz\\ 
Keyword(s): Lower dimensional models, canonical formalism, canonical
quantization
\end{abstract}
\newpage\noindent
\section{Introduction}
Recently, a lot of attention has been devoted to the investigation of
lower-dimensional gravity \cite{web}. The interest on dimensionally
reduced theories of gravity relies essentially on their connection to
string theory, black hole physics, and gravitational collapse. In this
context, two-dimensional models of dilaton gravity play a very important
role because of their relation to higher-dimensional gravity and
integrable systems.

Two-dimensional dilaton-gravity is described by the action 
\be
S_{DGM}=\int_\Sigma d^2x\sqrt{-\gamma}\,[{\cal U}(\phi){\cal
R}^{(2)}(\gamma)+{\cal V}(\phi)+ {\cal 
W}(\phi)(\nabla\phi)^2]+S_M[\phi,\gamma_{\mu\nu},f_i]\,,\label{action-gen}
\ee
where $\cal U$, $\cal V$, and $\cal W$ are arbitrary functions of the
dilaton, ${\cal R}^{(2)}$ is the two-dimensional Ricci scalar, and $S_M$
represents the contribution of matter fields $f_i$ which include any field
but the dilaton $\phi$ and the graviton $\gamma_{\mu\nu}$.

Most of the models studied in some detail in the literature are special
cases of the model described by Eq.\ (\ref{action-gen}) where dilaton
gravity is coupled to scalar, gauge, and fermion fields. (See for instance
Refs.\ \cite{Filippov}-\cite{CavaPLB} and references therein.) For a given
$S_M$, Eq.\ (\ref{action-gen}) describes a family of models whose elements
are identified by the choice of the dilatonic potential. Indeed,
classically we may always choose ${\cal U}(\phi)=\phi$ and locally set
${\cal W}(\phi)=0$ by a Weyl-rescaling of the metric. (In this paper we
will always make this choice for simplicity.) So the matterless sector of
Eq.\ (\ref{action-gen}) reads
\be
S_{DG}=\int_\Sigma d^2x\sqrt{-g}\,[\phi R^{(2)}(g)+V(\phi)]\,,
\label{action-mless}
\ee
where $g_{\mu\nu}$ and $R^{(2)}$ are the two-dimensional, Weyl-rescaled,
metric and Ricci scalar respectively. Different choices of $V(\phi)$
identify different theories. Remarkable examples are the
Callan-Giddings-Harvey-Strominger (CGHS) model \cite{CGHS} ($V=const$),
the Jackiw-Teitelboim model \cite{JT} ($V=\phi$), and the dimensionally
reduced theory of four-dimensional spherically-symmetric Einstein gravity
integrated on the two-sphere of area $16\pi\phi$ ($V=1/(2\sqrt{\phi})$)
\cite{LGK}.

According to their integrability properties, dilaton gravity models can be
roughly divided in three classes: 
\begin{itemize}
\item {\it i)} {\it Completely Integrable Models}, i.e.\ models that can
be expressed in terms of free fields by a canonical transformation.
Remarkable examples are matterless dilaton gravity with an arbitrary
potential \cite{Filippov} and the CGHS model
\cite{Jetal,Kucetal}; 
\item {\it ii)} {\it Completely Solvable Models}, i.e.\
models that cannot be analytically solved in terms of free fields but
whose general solution is known. Two-dimensional effective generalized
theory of 2+1 cylindrical gravity minimally coupled to a massless scalar
field \cite{CavaPRD} and dilaton gravity with constant or linear dilatonic
potential minimally coupled to massless Dirac fermions \cite{CFF,PS}
belong to this class; 
\item {\it iii)} {\it Partially Integrable Models}, i.e.\ models
that are integrable in a 0+1 dimensional sector only, namely after
reduction to a finite number of degrees of freedom. In this category we
find, for example, dilaton gravity minimally coupled to massless Dirac
fermions with arbitrary potential \cite{CFF} and two-dimensional effective
models describing uncharged black $p$-branes in $N$ dimensions 
\cite{CavaPLB}.
\end{itemize}
Completely integrable models are of particular interest from the quantum
point of view. In this case we are able to quantize the theory (in
the free-field representation) and, hopefully, to discuss quantization
subtleties and non-perturbative quantum effects. (See e.g.\ Refs.\
\cite{Jetal,Kucetal,Birkhoff} for the CGHS model.) In particular, matterless
dilaton gravity -- Eq.\ (\ref{action-mless}) -- can be used to describe
black holes and, in the case of coupling with scalar matter, gravitational
collapse. So the quantization program is worth exploring.

Although the classical properties of the model based upon Eq.\
(\ref{action-mless}) are well-known, not much is known about its
quantization: only the CGHS model has been investigated in depth. The two
most fruitful attempts to construct the quantum theory of the CGHS model
are described in Refs. \cite{Jetal,Kucetal,Birkhoff} and Ref.\
\cite{Varadarajan} respectively. 

The first approach is based on a canonical transformation mapping the original
system to a system described by free fields.  Then the theory is quantized in
the free field representation.  The main drawback of this approach is that the
new canonical variables are not directly related to the original spacetime
geometry and important physical quantities cannot be expressed in terms of the
new fields \cite{Birkhoff}.  Further, it is not clear how to generalize the
canonical transformation for an arbitrary dilatonic potential.  (Recently, a
proof of the existence of a canonical transformation that generalizes the
canonical transformation used in the CGHS case has been derived by Cruz and
Navarro-Salas, see Ref.\ \cite{Navarro}.  Even though it seems reasonable to
guess the existence of a canonical transformation in the general case, the
relation between the new fields and the original geometrical variables remains a
puzzle.)

The ``geometrodynamical approach'' was originally developed by Kucha\v{r}
for the canonical description of the Schwarzschild black hole
\cite{Kuchar}. This approach uses variables that are directly related to
the spacetime geometry and does not make use of the field redefinitions of
Refs. \cite{Jetal,Kucetal,Birkhoff}.  Again, only the CGHS model has been
quantized using this formalism \cite{Varadarajan}.

In this paper we quantize the {\em general} matterless dilaton gravity
model described by Eq.\ (\ref{action-mless}) using a transformation of the
configuration space performed at the Lagrangian level. The transformation
is suggested by the topological nature of two-dimensional gravity and by
the existence of a local integral of motion independent of the coordinates
first discussed by Filippov \cite{Filippov}. The new fields have
clear physical meaning -- they are the dilaton and the ``mass'' of the
system -- thus avoiding problems related to their interpretation in terms
of the geometrical variables. 

In the canonical framework the new fields generalize the {\it
geometrodynamical variables} of Kucha\v{r} \cite{Kuchar} and Varadarajan
\cite{Varadarajan} to a generic dilatonic potential. Thus the quantization
is straightforward and can be completed along the lines of Refs.\
\cite{Kuchar,Varadarajan}.  The quantum theory reduces to quantum
mechanics and the Hilbert space coincides with the Hilbert space obtained
by quantizing the theory first reducing it to a 0+1 dynamical system with
a finite number of degrees of freedom and then imposing the quantization
algorithm (see Ref.\ \cite{bh} for the case $V=1/(2\sqrt{\phi}$)). This
result represents the quantum generalization of the well-known {\em
Birkhoff Theorem} for spherically symmetric gravity in four dimensions. (A
somewhat different derivation of the so-called {\em Quantum Birkhoff
Theorem} for the CGHS model is discussed in Ref.\ \cite{Birkhoff}. The
approach of Ref.\ \cite{Birkhoff} makes use of the canonical
transformation to free fields. Here we extend the results of Ref.\
\cite{Birkhoff} to the general model with arbitrary dilatonic potential
using a different and more powerful approach.)

The Quantum Birkhoff Theorem is schematically described by the following
diagram:
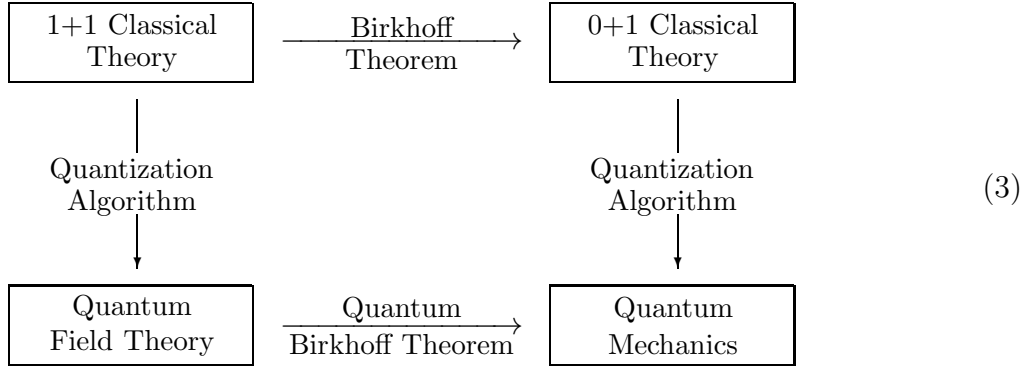
\be
\begin{array}{rcc}
\entry{1+1 Classical}{Theory}
&\arrwh{Birkhoff}{Theorem}&\entry{0+1 Classical}{Theory}\\
\arrwv{Quantization}{Algorithm}&&\arrwv{Quantization}{Algorithm}\\
\entry{Quantum}{Field Theory}&\arrwh{Quantum}{Birkhoff Theorem}&
\entry{Quantum}{Mechanics}
\end{array}
\ee

The outline of the paper is the following.  In the next section we briefly
review the classical theory of two-dimensional dilaton gravity and set up
notations. We follow essentially the approach developed by Filippov
\cite{Filippov}. In Sect.\ \ref{lagrangian} we introduce the point
transformation and the new Lagrangian.  In Sect.\ \ref{canonical} we discuss the
canonical framework.  Finally, in the last two sections we quantize the model
and state our conclusions.

\section{Classical Theory\label{classical}}

Let us consider Eq.\ (\ref{action-mless}). Varying the action w.r.t.\ 
the metric and the dilaton we obtain
\be
\bigl(\nabla_{(\mu}\nabla_{\nu)}-g_{\mu\nu}\nabla_\sigma\nabla^\sigma\bigr)
\phi+\displaystyle{1\over 2}g_{\mu\nu}V(\phi)=0\,,\label{eq-dil}
\ee
\be
R+\displaystyle{dV\over d\phi}=0\,,\label{eq-R} 
\ee 
where the symbol $\nabla$ represents covariant derivatives w.r.t.\ the
metric $g_{\mu\nu}$.

It is easy to prove that Eq.\ (\ref{eq-R}) is satisfied if Eq.\ (\ref{eq-dil}) 
is satisfied provided that
\be
H(g_{\mu\nu},\phi)\not=0\,,\label{horizon}
\ee
where $H(g_{\mu\nu},\phi)=\nabla_\rho\phi\nabla^\rho\phi$. This condition
can be lifted if one requires the continuity of the fields and of their
derivatives at any spacetime point. We will see in a moment -- see Eq.\
(\ref{horizon2}) below -- that the equation $H(g_{\mu\nu},\phi)=0$ defines
the horizon(s) of the two-dimensional metric. So by requiring the
continuity of the fields and their derivatives across the horizon(s) Eq.\
(\ref{eq-dil}) implies Eq.\ (\ref{eq-R}) everywhere.

The field equations (\ref{eq-dil}-\ref{eq-R}) can be solved performing a
B\"acklund transformation (see Ref.\ \cite{Filippov}). In covariant
language the B\"acklund transformation reads
\be
M=N(\phi)-\nabla_\rho\phi\nabla^\rho\phi\,,~~~~N(\phi)=\int^\phi
d\phi'V(\phi')\,,\label{transf-M}
\ee
\be
\nabla_\mu\psi=\displaystyle{\nabla_\mu\phi\over
\nabla_\rho\phi\nabla^\rho\phi}\,,\label{transf-psi}
\ee
where $M(t,x)$ and $\psi(t,x)$ are the transformed fields. Note
that the transformation is singular for $H(g_{\mu\nu},\phi)=0$. Using the
new fields Eq.\ (\ref{eq-dil}) reads
\be
\nabla_\mu\nabla^\mu\psi=0\,,\label{eq-psi}\\
\ee
\be
\nabla_\mu M=0\,.\label{eq-M}
\ee
Since the transformation (\ref{transf-M}-\ref{transf-psi}) is defined when
Eq.\ (\ref{horizon}) holds, Eqs.\ (\ref{eq-psi}-\ref{eq-M}) are equivalent
to the original field equations (\ref{eq-dil}-\ref{eq-R}) except at the
horizon(s). Equations (\ref{eq-psi}-\ref{eq-M}) have a deep significance. 
The first equation implies that $\psi$ is a free (D'Alembert) field. From
the second equation we find that $M$ is a locally conserved quantity. 

In two-dimensions any metric is locally conformally flat \cite{Eisenhart}.
So there exists a coordinate transformation which brings the metric into 
the form
\be
ds^2=4\rho(u,v)dudv\,,\label{metric-conf}
\ee
where $u=(t+x)/2$, $v=(t-x)/2$. Using conformal light-cone coordinates
Eqs.\ (\ref{eq-psi}-\ref{eq-M}) can be explicitly integrated. The general
solution is
\be
\psi=U(u)+V(v)\,,~~~~M=M_0\,.\label{cl-sol}
\ee
The original fields $\rho$ and $\phi$ are can be written as functions of
$\psi$ and $M$ using Eqs.\ (\ref{transf-M}-\ref{transf-psi}). With a
little algebra one finds
\be
{d\psi\over d\phi}=\displaystyle{1\over N(\phi)-M}\label{psi-phi}\,,
\ee
\be
\rho=[N(\phi)-M]\partial_u\psi\partial_v\psi\label{rho-psi}\,.
\ee
Equations (\ref{cl-sol}) and (\ref{psi-phi}-\ref{rho-psi}) imply that the
general solution of the model is actually (0+1)-dimensional, i.e.\ that
any solution possesses a Killing vector \cite{ks}. Indeed, using the
coordinates ($U,V$) the general solution reads
\be
ds^2=4[N(\phi)-M]dUdV\,,~~~~\phi\equiv\phi(U+V)\,,\label{cl-sol2}
\ee
or, using the coordinates ($\phi,T\equiv U-V$),
\be
ds^2=-[N(\phi)-M]dT^2+[N(\phi)-M]^{-1}d\phi^2\,.\label{cl-sol3}
\ee
Thus the general solution depends on the single variable $\phi$. (With a
somewhat improper terminology we call these solutions {\it static},
even though the Killing vector may not be timelike and hypersurface
orthogonal on the entire manifold.) This result constitutes a
generalization of the classical Birkhoff theorem \cite{Filippov,ks}. (For
spherically-symmetric Einstein gravity the ``local integral of motion
independent of the coordinates'' is just the Schwarzschild mass.) The
reduction of the theory to a finite-dimensional dynamical system signals
that pure dilaton gravity is actually a topological theory. In Sect.\
\ref{quantization} we will see how this property influences the
quantization of the theory.  

Let us briefly discuss the local geometrical properties of the solution
(\ref{cl-sol3}).  The horizon(s) of the metric are determined by the equation
\be
N(\phi)-M\equiv H(g_{\mu\nu},\phi)=0\,,\label{horizon2}
\ee

For a given choice of the dilatonic potential, Eq.\ (\ref{horizon2}) is an
algebraic equation in $\phi$ whose solutions $\{\phi_i\}$ determine the
values of the radial coordinate where the horizon(s) are located.  So the
request of continuity of the solution -- and of its derivatives, see Ref.\
\cite{Kuchar} -- across the horizons enforces the continuity of the fields
$\rho$ and $\phi$ at the points $H(g_{\mu\nu},\phi)=0$ and viceversa. This
justifies a posteriori the assumption of continuity made below Eq.\
(\ref{horizon}). With this assumption Eqs.\ (\ref{eq-dil}-\ref{eq-R}) are
equivalent to Eqs.\ (\ref{eq-psi}-\ref{eq-M}) everywhere. 

The local asymptotic structure of the solution (\ref{cl-sol3}) and the
existence of singularities depend on the choice of the dilatonic
potential. In particular, from Eq.\ (\ref{eq-R}) one finds that
singularities of the metric are determined by singular points of the first
derivative of $V(\phi)$ w.r.t. $\phi$. The local asymptotic structure can
be also roughly investigated using Eq.\ (\ref{eq-R}). For instance, let us
suppose that the asymptotic region is defined by $\phi\to\infty$ and that
the behavior of the dilatonic potential at infinity is $V(\phi)\approx
\phi^k$, where $k$ is a constant parameter. Thus the two-dimensional
spacetime is asymptotically flat for $\phi\to\infty$ if $k<1$, and has
constant curvature for $\phi\to\infty$ if $k=1$. 

Let us conclude this section with a concrete example and derive the
Schwarzschild solution using the formalism described above. The
dimensional reduction of the four-dimensional vacuum Einstein gravity
\be
S_{EH}={1\over 16\pi}\int_\Sigma
d^4x\sqrt{-g}\,R^{(4)}(g)\,,\label{action-EH}
\ee
can be obtained using the ansatz
\be
ds_{(4)}^2={1\over\sqrt{\phi}}g_{\mu\nu}dx^\mu
dx^\nu+4\phi\, d\Omega^2\,,~~~~~\phi\ge 0\,,\label{4dmetric}
\ee
where $g_{\mu\nu}$ is a two-dimensional metric with signature 
$(-1,1)$ and $d\Omega^2$ is the line element of the unit two-sphere.
Using Eq.\ (\ref{4dmetric}), and integrating on the two-sphere, the
four-dimensional Einstein-Hilbert action can be cast into the form
(\ref{action-mless}) with $V(\phi)=1/(2\sqrt{\phi})$. Using Eq.\
(\ref{cl-sol3}) the line element (\ref{4dmetric}) reads
\be
ds_{(4)}^2=-\left(1-{M\over\sqrt{\phi}}\right)dT^2+{d\phi^2\over\phi
\left(1-\displaystyle{M\over\sqrt{\phi}}\right)}+4\phi\,
d\Omega^2_2\,.\label{schw}
\ee
Clearly Eq.\ (\ref{schw}) reduces to the standard Schwarzschild
solution with the substitution $4\phi=R^2$.

\section{Lagrangian Formalism\label{lagrangian}}
The B\"acklund transformation introduced in the previous section can be
used to find a transformation from the original fields,
$(g_{\mu\nu},\phi)$, to new fields $(X_i)$, $i=1..4$, where one of the new
fields $X_i$ coincides with $M$. Since $M$ is a locally conserved quantity
this transformation simplifies drastically the dilaton gravity Lagrangian
in Eq.\ (\ref{action-mless}). 

The key of the construction is the observation that in two-dimensions
the Ricci scalar $R$ is a total divergence and can be locally written
as
\be
{R\over 2}=\nabla_\mu A^\mu\,,~~~~A^\mu=
{\nabla^\mu\nabla^\nu\chi\nabla_\nu\chi-
\nabla_\nu\nabla^\nu\chi\nabla^\mu\chi\over
\nabla_\rho\chi\nabla^\rho\chi}\,,\label{ricci-div}
\ee
where $\chi$ is an arbitrary, non-constant, function of the coordinates. 
Equation (\ref{ricci-div}) can be easily checked using conformal
coordinates. Since Eq.\ (\ref{ricci-div}) is a generally covariant
expression, and any two-dimensional metric can be locally cast in the form
(\ref{metric-conf}) by a coordinate transformation \cite{Eisenhart}, Eq.\
(\ref{ricci-div}) is valid in any system of coordinates.

Differentiating Eq.\ (\ref{transf-M}), and choosing $\chi=\phi$, both
$V(\phi)$ and $R$ can be written as functions of $M$ and $\nabla_\mu\phi$.
Finally, by an integration per parts we find
\be
S=\int_\Sigma d^2x\,\sqrt{-g}\,{\nabla_\mu\phi\nabla^\mu M
\over N(\phi)-M}+S_{\partial}\,,\label{action-new}
\ee
where $S_{\partial}$ is the surface term
\be
S_{\partial}=2\int_\Sigma 
d^2x\,\sqrt{-g}\,\nabla_\mu\left[\nabla^\mu\phi+\phi
A^\mu\right]\,.\label{surface}
\ee
Let us check that Eq.\ (\ref{action-new}) has the same number of d.o.f.\
of the original action (\ref{action-mless}). In two dimensions a generic
metric can be written
\be
g_{\mu\nu}=\rho\left(\matrix{\alpha^2-\beta^2&\beta\cr
\beta&-1\cr}\right)\,.\label{ADM}
\ee
In the canonical formalism $\alpha(t,x)$ and $\beta(t,x)$ play the role of
the lapse function and of the shift vector respectively;  $\rho(t,x)$ is
the dynamical d.o.f.. Due to the chosen parametrization, the Lagrangian in
Eq.\ (\ref{action-mless}) is a functional of the two dynamical fields
$(\rho,\phi$) and of the two non-dynamical variables ($\alpha,\beta$). Now
let us use Eq.\ (\ref{ADM}) in Eq.\ (\ref{action-new}) and neglect the
surface term. The new Lagrangian is again a functional of two fields
($M,\phi$) and of two non-dynamical variables ($\alpha,\beta$). Indeed,
since Eq.\ (\ref{action-new}) only contains the Weyl-invariant
combinations $\sqrt{-g}g^{\mu\nu}$, the transformed action is invariant
under changes of coordinates which belong to the conformal group and
$g_{\mu\nu}$ does not contribute any dynamical d.o.f.\ to the action. As a
consequence, the transformation
$(\rho,\phi,\alpha,\beta)\to(M,\phi,\alpha,\beta)$ is a ``point
transformation'' with $M\equiv M(\rho,\phi,\alpha,\beta)$ defined by
Eq.\ (\ref{transf-M}). (Quotation marks are due to the fact that the
transformation $(\rho,\phi,\alpha,\beta)\to(M,\phi,\alpha,\beta)$ should
not be regarded as a point transformation according to the usual lore
because it involves derivatives w.r.t.\ $t$ and $x$. We call it ``point
transformation'' because it can be implemented at the Lagrangian level.)

Varying Eq.\ (\ref{action-new}) we find
\be
\nabla_\mu\nabla^\mu\phi-V(\phi)=0\,,\label{eq-M2}
\ee
\be
\nabla_{(\mu}\phi\nabla_{\nu)}M-{1\over
2}g_{\mu\nu}\nabla_\sigma\phi\nabla^\sigma M=0\,,\label{eq-g}
\ee
\be
\nabla_\mu M\nabla^\mu
M+\nabla_\nu\phi\nabla^\nu\phi\nabla_\mu\nabla^\mu M=0\,.\label{eq-phi}
\ee
Equations (\ref{eq-M2}-\ref{eq-phi}) are equivalent to the field equations
obtained from Eq.\ (\ref{action-mless}). Equation (\ref{eq-M2}) 
corresponds to the trace of Eq.\ (\ref{eq-dil}). Further, by
differentiation of Eq.\ (\ref{transf-M}) one finds that Eqs.\
(\ref{eq-g}-\ref{eq-phi}) are satisfied if Eq.\ (\ref{eq-dil}) is
satisfied because Eq.\ (\ref{eq-dil}) implies $\nabla_\mu M=0$. The
converse latter statement is also true provided that Eq.\ (\ref{horizon}) 
is satisfied. When this condition holds Eq.\ (\ref{eq-g}) implies
$\nabla_\mu M=0$. By requiring the continuity of the fields Eqs.\
(\ref{eq-M2}-\ref{eq-phi}) and Eqs.\ (\ref{eq-dil}-\ref{eq-R}) are
equivalent. The equivalence of Eqs.\ (\ref{eq-M2}-\ref{eq-phi}) and the
original field equations can also be directly checked using the metric
parametrization defined in Eq.\ (\ref{ADM}). 

\section{Canonical Formalism\label{canonical}}

The canonical formalism is an essential step in the quantization procedure. 
Starting from Eq.\ (\ref{action-mless}), and using the metric
parametrization Eq.\ (\ref{ADM}), the action can be cast in the
Hamiltonian form
\be
S=\int 
dt\,\int_{x_a}^{x_b}dx\,\left[\pi_\rho\dot\rho+\pi_\phi\dot\phi-\alpha{\cal
H}_0-\beta{\cal H}_1\right]\,,\label{action-ham}
\ee
where dots represent derivatives w.r.t.\ the timelike coordinate $t$,
($\rho,\phi,\pi_\rho,\pi_\phi)$ are the phase space variables, and ${\cal
H}_0$, ${\cal H}_1$ are the ADM super-Hamiltonian and super-momentum
respectively:
\ba
&&{\cal
H}_0=\rho\pi_\rho\pi_\phi+\displaystyle{\rho'\over\rho}\phi'-2\phi''-\rho
V(\phi)\,,\\\\
&&{\cal H}_1=-\phi'\pi_\phi+\rho'\pi_\rho+2\rho\pi_\rho'\,.
\label{shm}
\ea
Here primes represent derivatives w.r.t.\ the spatial coordinate $x$. 
Equations (\ref{shm}) include, as particular cases, the models
discussed in Ref.\ \cite{Varadarajan} and Ref.\ \cite{Kuchar}.
Denoting with subscripts $v$ and $k$ the canonical variables of Ref.\
\cite{Varadarajan} and \cite{Kuchar} respectively, we have
\ba
&\phi=\displaystyle{R_v^2\over 4}\,,
&\pi_\phi=2\displaystyle{R_v P_{R_v}-\Lambda_v P_{\Lambda_v}\over
R_v^2}\,,\\\\
&\rho=R_v^2\Lambda_v^2\,,
&\pi_\rho=\displaystyle{P_{\Lambda_v}\over 2R_v^2
\Lambda_v}\,,\\\\
&\alpha=-\displaystyle{N_v\over\Lambda_v}\,,
&\beta=-N^r_v\,,
\label{m-v}
\ea
for the CGHS model ($V=const$), and
\ba
&\phi=\displaystyle{R_k^2\over 4}\,,
&\pi_\phi=\displaystyle{2R_k P_{R_k}-\Lambda_k P_{\Lambda_k}\over
R_k^2}\,,\\\\
&\rho=\displaystyle{R_k \Lambda_k^2\over 2}\,,
&\pi_\rho=\displaystyle{P_{\Lambda_k}\over R_k \Lambda_k}\,,\\\\
&\alpha=-\displaystyle{N_k\over\Lambda_k}\,,
&\beta=-N^r_k\,,
\label{m-k}
\ea
for the Schwarzschild black hole $V=-1/(2\sqrt{\phi})$. (The minus sign of
$V$ is due to the choice of the metric signature in Eq.\ (\ref{ADM}) that
is opposite to the signature used in Eq.\ (\ref{4dmetric}).) 

Starting from Eq.\ (\ref{action-new}) the super-Hamiltonian and
super-momentum read (for later convenience we set $\phi=\bar\phi$)
\ba
&&{\cal H}_0=[N(\bar\phi)-M]\pi_{\bar\phi}\pi_M+
[N(\bar\phi)-M]^{-1}\bar\phi'M'\,,\\\\
&&{\cal H}_1=-\bar\phi'\pi_{\bar\phi}-M'\pi_M\,.
\label{smh2}
\ea
Eventually, both canonical actions must be complemented by a
boundary term at the spatial boundaries. This can be done along the lines
of Refs.\ \cite{Kuchar,Varadarajan} as we will see later in this section.

The two charts $(\phi,\pi_{\phi},\rho,\pi_\rho)$ and  
$(\bar\phi,\pi_{\bar\phi},M,\pi_M)$ are related by the transformation
\ba
&&M=N(\phi)-\displaystyle{\rho^2\pi_\rho^2-\phi'^2\over\rho}\,,\\
&&\pi_M=\displaystyle{\rho^2\pi_\rho\over 
\rho^2\pi_\rho^2-\phi'^2}\,,\\
&&\bar\phi=\phi\,,\\
&&\pi_{\bar\phi}=\pi_\phi-\displaystyle{\rho^2\pi_\rho\over
\rho^2\pi_\rho^2-\phi'^2}\left[V(\phi)+2\pi_\rho\left({\phi'\over
\rho\pi_\rho}\right)'\right]\,.
\label{tr}
\ea
The transformation given above is easily invertible. The result is:
\ba
&&\rho=\pi_M^2\bigl(N(\bar\phi)-M\bigr)\left[1-\left(
\displaystyle{\bar\phi'\over\pi_M
\bigl(N(\bar\phi)-M\bigr)}\right)^2\right]\,,\\\\
&&\pi_\rho=\displaystyle{1\over
\pi_M\left[1-\left(
\displaystyle{\bar\phi'\over\pi_M
\bigl(N(\bar\phi)-M\bigr)}\right)^2\right]}\,,\\\\
&&\phi=\bar\phi\,,\\\\
&&\pi_\phi=\pi_{\bar\phi}+V(\bar\phi)\pi_M+\displaystyle{2\left(
\displaystyle{\bar\phi'\over\pi_M\bigl(N(\bar\phi)-M\bigr)}\right)'\over
\left[1-\left(\displaystyle{\bar\phi'\over\pi_M
\bigl(N(\bar\phi)-M\bigr)}\right)^2
\right]}\,.
\label{tr-inv}
\ea
After some tedious calculations one can check that the only non-vanishing
Poisson brackets at equal time $t$ are
\be
\bigl[M(t,x),\pi_M(t,x')\bigr]=\delta(x-x')\,,~~~~~
\bigl[\bar\phi(t,x),\pi_{\bar\phi}(t,x')\bigr]=\delta(x-x')\,,
\ee
so Eqs.\ (\ref{tr}-\ref{tr-inv}) define a canonical map. Finally, the 
difference of the Liouville forms reads
\be
\int_{x_a}^{x_b}dx\,(\dot M\,\pi_M+\dot{\bar\phi}\,\pi_{\bar\phi})-
\int_{x_a}^{x_b}dx\,(\dot\rho\,\pi_\rho+\dot\phi\,\pi_\phi)=
F(\bar\phi,\pi_{\bar\phi},M,\pi_M)\,,\label{diff-liouville}
\ee
where
\ba
&&
F(\bar\phi,\pi_{\bar\phi},M,\pi_M)=\displaystyle\int_{x_a}^{x_b}dx\,
\left\{2\bar\phi'\,{\rm arctanh}\,\left[\displaystyle{\bar\phi'\over
\bigl(N(\bar\phi)-M\bigr)\pi_M}\right]-2\bigl(N(\bar\phi)-M\bigr)
\pi_M\right\}^{\bf\cdot}+\\\\
&&\hbox to 170truept{}
-\displaystyle\int_{x_a}^{x_b}dx\,\left\{2\dot{\bar\phi}\,{\rm arctanh}\,
\left[\displaystyle{\bar\phi'\over
\bigl(N(\bar\phi)-M\bigr)\pi_M}\right]\right\}'\,.\label{F}
\ea

The canonical variables $(\bar\phi,\pi_{\bar\phi},M,\pi_M)$ are a
generalization of the geometrodynamical variables introduced by Kucha\v{r}
\cite{Kuchar} and Varadarajan \cite{Varadarajan}. This can be easily
proved using Eqs.\ (\ref{m-v}-\ref{m-k}) and Eqs.\ (\ref{tr}-\ref{tr-inv}). 

Now we must take care of boundary terms and define falloff conditions
at the spatial boundaries. We set
\ba
&&\bar\phi'=[N(\bar\phi)-M]\bigl(1+\epsilon^{(a,b)}_{\bar\phi'}\bigr)\,,\\\\
&&\dot{\bar\phi}=\epsilon^{(a,b)}_{\dot{\bar\phi}}\,,\\\\
&&M=M^{(a,b)}(t)\bigl(1+\epsilon^{(a,b)}_M\bigr)\,,\\\\
&&\pi_{\bar\phi}=\displaystyle{\epsilon^{(a,b)}_{\pi_{\bar\phi}}\over
N(\bar\phi)-M}\,,\\
&&\pi_M=\displaystyle{\epsilon^{(a,b)}_{\pi_M}\over
N(\bar\phi)-M}\,,\\\\
&&\alpha=\alpha^{(a,b)}(t)\bigl(1+\epsilon^{(a,b)}_\alpha\bigr)\,,\\\\
&&\beta=\epsilon^{(a,b)}_\beta\,,
\label{falloff}
\ea
where $\epsilon^{(a,b)}$ are functions of $t$ and $x$ vanishing at
the spatial boundaries $x_a$ and $x_b$, i.e.\
\be
\lim_{x\to x_a,x_b}\epsilon^{(a,b)}(t,x)=0\,,
\ee
and 
\ba
&&\displaystyle\lim_{x\to 
x_a,x_b}\displaystyle{\epsilon^{(a,b)}_{\pi_{\bar\phi}}\over
N(\bar\phi)-M}=0\,,\\\\
&&\displaystyle\lim_{x\to x_a,x_b}\displaystyle{{\epsilon^{(a,b)}_{\pi_M}}\over
N(\bar\phi)-M}=0\,,\\\\
&&\displaystyle\lim_{x\to x_a,x_b}\displaystyle{{\epsilon^{(a,b)}_M}'\over
N(\bar\phi)-M}=0\,.
\ea
The exact behavior of the $\epsilon^{(a,b)}$ functions depends on the
particular potential under consideration. By requiring that
$\epsilon^{(a,b)}$ go to zero rapidly enough, both the Liouville form and
the super-Hamiltonian and super-momentum are well defined and the
difference of the Liouville forms $F(\bar\phi,\pi_{\bar\phi},M,\pi_M)$ reduces 
to an exact form. For instance, in the Schwarzschild black hole case
$N(\phi)=\sqrt{\phi}$ the spatial boundaries are located at $x_a=-\infty$ and
$x_b=+\infty$ and with a little algebra one can check that the falloff
conditions (\ref{falloff}) become
\ba
&&\bar\phi=\displaystyle{x^2\over
4}\bigl(1+\bar\phi_\pm 
|x|^{-1-\epsilon}+O^{\pm\infty}(|x|^{-2-\epsilon})\bigr)\,,\\\\
&&M=M_{\pm}(t)\bigl(1+O^{\pm\infty}(|x|^{-\epsilon}\bigr)\,,\\\\
&&\pi_{\bar\phi}=O^{\pm\infty}(|x|^{-2-\epsilon})\,,\\\\
&&\pi_M=O^{\pm\infty}(|x|^{-1-\epsilon})\,,\\\\
&&\alpha=\alpha_\pm (t)
\bigl(1+O^{\pm\infty}(|x|^{-\epsilon})\bigr)\,,\\\\
&&\beta=O^{\pm\infty}(|x|^{-\epsilon})\,,~~~~~~0<\epsilon\le 1\,.
\label{Kuchar-falloff}
\ea
In this case the super-Hamiltonian and super-momentum fall off as
\be
{\cal H}_0=O^{\pm\infty}(|x|^{-1-\epsilon})\,,~~~~~{\cal
H}_1=O^{\pm\infty}(|x|^{-1-\epsilon})\,,
\ee
and the Liouville form
\be
\int_{-\infty}^{+\infty}dx\,(\dot
M\pi_M+\dot{\bar\phi}\pi_{\bar\phi})\,,~~~~~
\dot M\pi_M+\dot{\bar\phi}\pi_{\bar\phi}=O^{\pm\infty}(|x|^{-1-\epsilon})
\ee
is well-defined. Finally, using Eqs.\ (\ref{Kuchar-falloff}) the second
term in the r.h.s.\ of Eq.\ (\ref{F}) vanishes at spatial infinities and
the difference of the Liouville forms is an exact form. The above falloff 
conditions coincide with those used by Kucha\v{r} in Ref.\ \cite{Kuchar}.

Following \cite{RT} we must complement the action by a boundary term to
allow functional differentiability of the action. Using the falloff
conditions (\ref{falloff}) we have that the sole non-vanishing boundary
term due to the variation of the action w.r.t.\ the canonical variables
has the form
\be
\int dt\,\bigl(-\alpha^{(b)}(t)\delta M^{(b)}+\alpha^{(a)}(t)\delta
M^{(a)}\bigr)\,,
\ee
so we add to the action the boundary term
\be
S_{\rm boundary}=\int
dt\,\bigl(\alpha^{(b)}(t)M^{(b)}(t)-\alpha^{(a)}(t)M^{(a)}(t)
\bigr)\,,\label{boundary}
\ee
where $\alpha^{(a,b)}(t)$ parametrize the action at the boundary and are
interpreted as prescribed values of $t$ (see Ref.\ \cite{Kuchar} for a
more detailed discussion about this point).

The canonical field equations and the constraints ${\cal H}_0=0$, ${\cal
H}_1=0$ are easily solved using the geometrodynamical chart
$(\bar\phi,\pi_{\bar\phi},M,\pi_M)$. The general solution of the
constraints is
\be
\pi_{\bar\phi}=0\,,~~~~~M'=0\,.\label{sol-constr}
\ee
Equations (\ref{sol-constr}) have the same physical content of
Eq.\ (\ref{eq-M}). (Note that $M$ weakly commutes with the constraints, as
expected for a local integral of motion.) Equation (\ref{eq-psi}) is a
direct consequence of the canonical field equations.

\section{Quantization\label{quantization}}
The quantization of the full 1+1 theory can be implemented using the
geometrodynamical canonical variables. From Eqs.\ (\ref{sol-constr}) we
read that $M$ does not depend on the spacelike coordinate $x$. The
effective Hamiltonian is simply given by the boundary term
(\ref{boundary}) and the reduced action reads
\be
S_{\rm eff}=\int d\tau\left[\frac{dm}{d\tau} 
p_m-m\right]\,,\label{action-red}
\ee
where $m\equiv M^{(b)}(t)=M^{(a)}(t)$,
$p_m\equiv\int_{x_a}^{x_b}dx\,\pi_M$ and $\tau(t)=\int^{t}
dt'(\alpha^{(a)}(t')-\alpha^{(b)}(t'))$. The theory reduces formally to
quantum mechanics and the quantization can be carried on as usual. The
Schr\"odinger equation is
\be
i{\d\over\d\tau}\Psi(m;\tau)={\cal H}_{\rm eff}~\Psi(m;\tau)\,,
~~~~~~~~~{\cal H}_{\rm eff}\equiv m\,.
\ee
The stationary states are the eigenfunctions of $m$ and the Hilbert space
coincides with the Hilbert space obtained in the 0+1 approach. Let us see
briefly this point in detail.

In the 0+1 approach we take advantage that every solution is static --
according to the definition given below Eq.\ (\ref{cl-sol3}) -- and set from 
the beginning $g_{\mu\nu}\equiv g_{\mu\nu}(t)$, $\phi\equiv\phi(t)$.  The
action (density) reads 
\be
S_{0+1}=\int dt\,[\dot\rho\pi_\rho+\dot\phi\pi_\phi-\alpha{\cal H}]\,,
\ee
where $\alpha$ is a Lagrange multiplier enforcing the constraint
${\cal H}=0$. ($\cal H$ corresponds to the 0+1 slice of ${\cal H}_0$. The
super-momentum constraint ${\cal H}_1$ vanishes identically.)

So in the 0+1 sector of the theory we can express the field equations as a
canonical system in a finite, $2\times 2$ dimensional, phase space.  The
equations of motion are analytically integrable and their solution
coincides with the finite gauge transformation generated by the constraint
${\cal H}=0$. We can find a couple of gauge-invariant canonically
conjugate quantities, $m$ and $p_m$, corresponding to the $0+1$ sections
of $M$ and $\pi_M$ introduced in Eqs.\ (\ref{tr}). The canonical variables
$m$ and $p_m$ can be identified with the quantities defined below Eq.\
(\ref{action-red}). 

Now we can construct the maximal gauge-invariant canonical chart
$(m,p_m,{\cal H},{\cal T})$ and use $\cal T$ to fix the gauge.  Indeed,
the transformation properties of $\cal T$ under the gauge transformation
generated by $\cal H$ imply that time defined by this variable covers once
and only once the symplectic manifold, i.e.\ time defined by $\cal T$ is a
global time. The quantization becomes trivial and the Hilbert space is
spanned by the eigenvectors of the sole -- apart from its conjugate
momentum -- gauge invariant operator $m$ corresponding to the mass of the
system. 

This quantization program has been implemented in detail in Refs.\
\cite{bh} for the case of spherically-symmetric Einstein gravity but can
be easily generalized to an arbitrary $V(\phi)$. (See for instance Ref.\
\cite{vda-dubna}.) In the 0+1 approach one can go further and discuss the
self-adjointness properties of the mass operator. It turns out that the
Hermitian operator $m$ in the gauge fixed, positive norm, Hilbert space is
not self-adjoint, while its square is a self-adjoint operator with
positive eigenvalues.  This result is due to the fact that the conjugate
variable to the mass, $p_m$, has positive support, analogously to what
happens for the radial momentum in ordinary quantum mechanics.  However,
the relevant point is that the mass $m$ -- or its square -- is the only
gauge-invariant observable of the system (apart from the conjugate
variable, of course) and the Hilbert space of the 0+1 approach coincides
with the Hilbert space of the full quantum 1+1 theory obtained through the
geometrodynamical formalism. This is the essence of the quantum Birkhoff
theorem.

\section{Conclusions and Perspectives}
Let us conclude with few remarks. We have derived a canonical
transformation to geometrodynamical variables that generalizes the
transformation of Ref.\ \cite{Varadarajan} and Ref.\ \cite{Kuchar} to any
dilaton gravity model. We have seen that the general dilaton gravity
action Eq.\ (\ref{action-mless}) can be cast into the form
(\ref{action-new}) and the system can be described both in the canonical
and Lagrangian frameworks using the dilaton and the mass as new variables.
The quantization of the general dilaton gravity model becomes
straightforward and the resulting quantum theory exhibits at the quantum
level the Birkhoff theorem. 

We believe that Eq.\ (\ref{action-new}) and Eqs.\ (\ref{smh2}) can be used
to look at two-dimensional gravity from a new perspective. Up to now
people have struggled themselves to find a canonical transformation
mapping the general dilaton gravity theory based upon Eq.\
(\ref{action-mless}) into a system described by free fields -- see for
instance Ref.\ \cite{Navarro}. Even though it seems reasonable to assume
the existence of such canonical transformation, we know from the CGHS case
(the simplest possible case!) that the relation between the free fields
and the ``physical'' fields (metric, dilaton, mass) is highly non-linear.
Further, the canonical transformation may be patological and subtleties
and ambiguities may arise. For instance, in the CGHS case the gauge
invariant operator $M$ cannot be expressed as a function of the new free
fields, as we would expect implementing a canonical transformation
\cite{Birkhoff}. This means that the canonical transformation to free
fields is ill-defined. Indeed, a careful analysis shows that the
transformation cannot be inverted and, in order to make it invertible, one
has to supplement the new field variables by an extra pair of conjugate
variables related to the value of the fields at the boundary. (See for
instance \cite{Birkhoff} and \cite{Kucetal}.) In spite of this
difficulties, the CGHS model can still be managed and different approaches
to the quantization can be carried on, leading to a consistent quantum
theory \cite{Jetal,Kucetal,Birkhoff}. However, we find very hard to
believe that models with more complicated dilatonic potentials can be
dealt with using free fields. Eventually, one wants quantum
operators corresponding to the physical quantities of the model i.e.,
quoting Kucha\v{r}, Romano, and Varadarajan, ``..the interesting questions
in dilatonic gravity are precisely those which are concerned with the
physical spacetime..'' \cite{Kucetal}. Free fields are very distant from
this picture. 

Conversely, the canonical variables defined in Eq.\ (\ref{tr}), being
directly related to the spacetime geometry, do not suffer from the
problems outlined above. Thus a quantum theory in which quantum operators
have a clear physical meaning is easily achieved. Finally, Eq.\
(\ref{action-new}) may (hopefully) provide a completely new starting point
in the investigation of open issues as, for instance, thermodynamics of
black holes and gravitational collapse (when matter is coupled to the
system).

\section*{Acknowledgements}
I am indebted to Vittorio de Alfaro, Alexandre T.\ Filippov, Claus Kiefer,
and Jorma Louko for many interesting discussions and useful suggestions
about the subject of this paper.  This work has been supported by a Human
Capital and Mobility grant of the European Union, contract no.\
ERBFMRX-CT96-0012. 

\end{document}